\documentclass[aps,prl,floats,floatfix,twocolumn,showpacs,superscriptaddress]{revtex4}
\usepackage{latexsym,amsmath}
\usepackage{amsbsy}
\usepackage[pdftex]{graphicx}
\usepackage{amssymb}
\usepackage{epstopdf}
\usepackage[usenames]{color}

\begin{document}

\title{Percolation in self-similar networks}

\author{M. \'Angeles Serrano}
\affiliation{Departament de Qu\'imica F\'isica, Universitat de Barcelona, Mart\'i i Franqu\`es 1, 08028, Barcelona, Spain}

\author{Dmitri Krioukov}

\affiliation{Cooperative Association for Internet Data
Analysis (CAIDA), University of California, San Diego (UCSD), 9500
Gilman Drive, La Jolla, CA 92093, USA}

\author{Mari{\'a}n Bogu{\~n}{\'a}}

\affiliation{Departament de F{\'\i}sica Fonamental, Universitat de
  Barcelona, Mart\'{\i} i Franqu\`es 1, 08028 Barcelona, Spain}

\date{\today}

\begin{abstract}

We provide a simple proof that graphs in a general class of self-similar networks have zero percolation threshold. The considered self-similar networks include random scale-free graphs with given expected node degrees and zero clustering, scale-free graphs with finite clustering and metric structure, growing scale-free networks, and many real networks. The proof and the derivation of the giant component size do not require the assumption that networks are treelike. Our results rely only on the observation that self-similar networks possess a hierarchy of nested subgraphs whose average degree grows with their depth in the hierarchy. We conjecture that this property is pivotal for percolation in networks.

\end{abstract}


\maketitle

Percolation is a fundamental phenomenon in nature. Recent developments in percolation theory~\cite{Smirnov:FieldsMedal} open new perspectives in many areas of statistical mechanics and quantum field theory~\cite{Cardy:ExplainsStuff}. In statistical mechanics of complex networks, the percolation properties of a network determine its robustness with respect to structural damage, and dictate how emergent phenomena depend on the network structure~\cite{Dorogovtsev:2008kx}.
Large clusters of connected nodes emerge above a critical value of some network parameter, e.g., the average degree; below the threshold, networks decompose into a myriad of small components. This percolation threshold can be zero, meaning that networks
are always in the percolated phase. A classic example is random scale-free networks with the power-law degree distribution exponent $\gamma$ lying between $2$ and $3$~\cite{Molloy:1995rh,Cohen:2000ss}. The value of the percolation threshold, the size of the giant component, and the specifics of the percolation transition strongly depend on fine details of the network topology~\cite{Dorogovtsev:2008kx}.
This dependency hinders attempts to define percolation universality classes, even though some networks show some degree of percolation universality~\cite{Krapivsky:2004bc}.

This problem is aggravated by difficulties in the analytic treatment of percolation properties for networks with strong clustering. A majority of the obtained analytic results use the generating function formalism based on the assumption that networks are locally treelike~\cite{Vazquez:2003lg}. This assumption allows one to employ convenient tools from the theory of random branching processes. The assumed absence of loops implies, in particular, that clustering is zero in the thermodynamic limit. This zero-clustering approximation is valid for weakly clustered networks where triangles do not overlap, but it is invalid for networks with strong clustering and overlapping triangles observed in many real systems~\cite{Serrano:2006qj}. Noticeably, the exact results derived for some network models with clustering can be mapped to treelike zero-clustering graphs after appropriate transformations~\cite{Newman:2003ft}.

In this Letter, we provide a remarkably simple rigorous proof for the absence of a percolation threshold in a general class of self-similar networks. The proof does not rely on the treelike assumption or on generating functions. It does not depend on whether a network is weakly or strongly clustered, and it applies equally well to equilibrium or non-equilibrium networks. The proof relies only on network self-similarity, defined as statistical invariance of a hierarchy of nested subgraphs with respect to a network renormalization procedure. The percolation threshold is zero as soon as the average degree in subgraphs is a growing function of their depth in the hierarchy---a property characterizing many real networks. We also calculate analytically the size of the giant component, supporting all the results by large-scale numerical simulations.

Let ${\cal G}(\{ \alpha \})$ be an ensemble of sparse graphs in the thermodynamic limit, where $\{ \alpha \}$ is the set of model parameters. In the case of classical random graphs, for example, set $\{ \alpha \}$ is just the average degree $\langle k \rangle$. Consider a transformation rule $T$ that for each graph $G \in {\cal G}(\{ \alpha \})$ selects one of $G$'s subgraphs. Denote the ensemble of these subgraphs by ${\cal G}_T(\{ \alpha \})$. The ensemble ${\cal G}(\{ \alpha \})$ is called self-similar with respect to the transformation rule $T$ if the transformed ensemble is the same as the original one except for some transformation of the model parameters,
\begin{equation}\label{eq:ss-def}
{\cal G}_T(\{ \alpha \})={\cal G}(\{ \alpha_T \}).
\end{equation}
In what follows we describe three general types of graphs to which this definition applies. The first two types are equilibrium random scale-free graph ensembles belonging to a general class of network models with hidden variables~\cite{Boguna:2003ja}. The third one is a non-equilibrium ensemble of growing networks.

{\bf Type I:} The graphs in this ensemble are constructed by assigning to each node a hidden variable $\kappa$ drawn from the power-law probability density $\rho(\kappa)=(\gamma-1) \kappa_0^{\gamma-1} \kappa^{-\gamma}$. Without loss of generality, $\kappa_0$ can be selected such that $\kappa\ge\kappa_0$ is the expected degree of nodes with hidden variable $\kappa$, so that the degree distribution scales as a power law with exponent $\gamma$. Each pair of nodes with expected degrees $\kappa$ and $\kappa'$ is then connected with probability
$
r(\kappa,\kappa')=f(\mu \kappa \kappa'),
\label{r_I}
$
where constant $\mu$ fixes the average degree $\langle k \rangle$ in the constructed graphs, and function $f(x) \le 1$ is an arbitrary analytic function with $f(0)=0$. This type of graphs includes as particular cases the maximally random graphs with a given expected degree sequence~\cite{Park:2003qy}, and random graphs with arbitrary structural correlations~\cite{Boguna:2004eh}. In the former case,
\begin{equation}
f(x)=\frac{1}{1+1/x}.
\label{r_typeI}
\end{equation}
Clustering vanishes in the thermodynamic limit, and therefore the treelike assumption holds.

{\bf Type II:} Besides having assigned expected degrees $\kappa$, nodes in this type of graphs are also uniformly distributed in a homogeneous and isotropic $D$-dimensional metric space~\cite{Serrano:2008ga}. Here, we consider a circle of radius $R$ with a constant density of nodes $\delta=N/(2\pi R)$, although all the following results can be extended to an arbitrary dimension. The connection probability between a pair of nodes with hidden variables $\kappa$ and $\kappa'$ separated by distance $d=(\pi-|\pi-|\theta-\theta'||)R$ on the circle ($\theta$'s are the node angular coordinates) must be of the form
$
r(\kappa,\theta;\kappa',\theta')=h\left(\frac{d}{\mu \kappa \kappa'}\right),
\label{r_II}
$
where function $h$ must be integrable. These graphs have the same degree distribution as the type I graphs, but clustering is finite in the thermodynamic limit, thanks to the triangle inequality in the underlying metric space~\cite{Serrano:2008ga}. Therefore the treelike assumption does not hold.

The graph sparsity in the thermodynamic limit defines constant $\mu$ in the two cases as
\begin{equation}
\mu_I=\frac{\langle k \rangle}{N f'(0) \kappa_0^2}\left(\frac{\gamma-2}{\gamma-1}\right)^2,\;
\mu_{II}=\frac{\langle k \rangle}{2 \delta I \kappa_0^2}\left(\frac{\gamma-2}{\gamma-1}\right)^2,
\end{equation}
where $I=\int_0^{\infty}h(x)dx$. Since $\kappa_0$ and $\delta$ are dumb parameters, we see that unless functions $f$ and $h$ contain some additional parameters, the described two graph ensembles have only two independent parameters: the power-law exponent $\gamma$ and the average degree $\langle k \rangle$.

Consider now transformation rule $T$ which simply removes all nodes with hidden variable $\kappa<\kappa_T$ from a given graph $G$ in any of the two ensembles, where $\kappa_T$ is some predefined threshold. This transformation maps the original graph $G$ to its subgraph $G_T$ of size $N_T=N (\kappa_0/\kappa_T)^{\gamma-1}$. The hidden variables $\kappa$ of nodes remaining in $G_T$ are distributed according to $\rho_T(\kappa)=(\gamma-1) \kappa_T^{\gamma-1} \kappa^{-\gamma}$ with $\kappa \ge \kappa_T$. That is, the power-law exponent in $G_T$ is the same as in $G$, $\gamma_T=\gamma$. The transformation does not affect the hidden variables of the nodes in subgraph $G_T$. Therefore the connection probability in $G_T$ is exactly the same as in the original graph $G$, which means that the ensemble of transformed graphs is identical to the ensemble of original graphs, except that the average degree has changed. Specifically, the transformation of parameters $\{\alpha\}\to\{\alpha_T\}$ in the self-similarity definition in Eq.~(\ref{eq:ss-def}) is
\begin{equation}
\gamma\to\gamma_T=\gamma,\quad
\langle k \rangle \to \langle k \rangle_T  = \langle k \rangle\left(\frac{N}{N_T}\right)^{\frac{3-\gamma}{\gamma-1}},
\label{scaling:1}
\end{equation}
which is the same for both type I and type II graphs~\cite{Serrano:2008ga}. Therefore, both ensembles belong to the same self-similarity universality class.

{\bf Type III:} As opposed to the first two equilibrium ensembles, the graphs of this type are grown by adding nodes one by one. Each node $i$ brings $m_i$ new links, where $m_i=m_0(N/i)^{\eta}$ and $\eta\in[0,1)$. Each link is then attached to a random existing node. A network is initialized with $10 m_0N^\eta$ uncounted disconnected nodes. If $\eta=0$, the generated graphs have an exponential degree distribution. If $\eta>0$, the degree distribution is $P(k) = (1+m_0/\eta)^{1/\eta}/[\eta(k+m_0/\eta)^{1+1/\eta}]$, i.e., approximately a power law with exponent $\gamma=1+1/\eta$. The transformation rule $T$ simply extracts from a grown graph its subgraph composed of the first $N_T$ nodes. With this $T$, the graphs of this type are also self-similar, and the parameter transformation in definition~(\ref{eq:ss-def}) is
\begin{equation}
\gamma\to\gamma_T=\gamma,\quad
\langle k \rangle \to \langle k \rangle_T = \langle k \rangle\left(\frac{N}{N_T}\right)^{\frac{1}{\gamma-1}}.
\label{scaling:2}
\end{equation}

\begin{figure}[t]
\centerline{\includegraphics[width=2.6in]{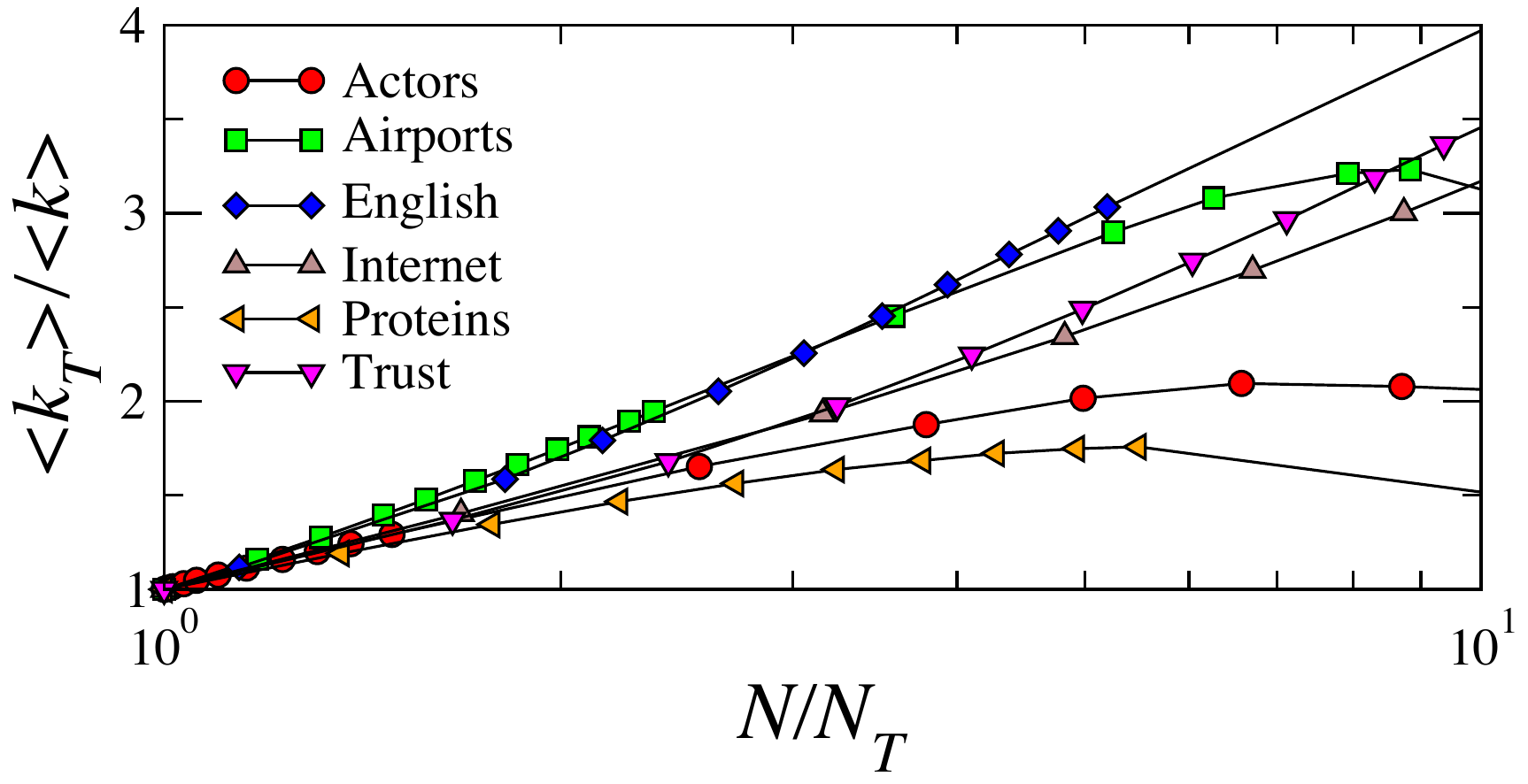}}
\caption{Ratio of the subgraph average degree $\langle k \rangle_T$ to the average degree $\langle k \rangle$ in the whole graph as a function of the inverse relative subgraph size $N/N_T$ for few real networks. The subgraphs are obtained by removing nodes with degrees below thresholds $k_T$ from the original network. To insulate against finite-size effects, the data is shown only for subgraphs of size $N_T/N>0.1$. {\it Actors}, actor collaborations from the Internet Movie Database; {\it Airports}, USA airport network; {\it English}, web of semantic associations between words in English; {\it Internet}, topology of the Internet at the Autonomous Systems level; {\it Proteins}, protein interaction network of {\it Saccharomyces cerevisiae}; and {\it Trust}, mutual trust relationships among individuals extracted from the Pretty Good Privacy data.}
\label{fig1}
\vspace{-0.5cm}
\end{figure}

As we show next, self-similarity of the considered ensembles (types I, II, and III), and the proportionality $N_T \sim N$ are sufficient to prove the absence of a percolation threshold for equilibrium scale-free graphs with exponent $\gamma<3$ and for growing graphs with any $\gamma$. The key property which we will use is that the average degree of these self-similar subgraphs is a growing function of the subgraph depth in the nested subgraph hierarchy, meaning that $\langle k \rangle_T$ grows as $N_T$ decreases in Eqs.~(\ref{scaling:1},\ref{scaling:2}). The same property characterizes many real networks as shown in Fig.~\ref{fig1} and in~\cite{Serrano:2008ga,Alvarez-Hamelin:2008fv}.

The proof is by contradiction. As usual~\cite{Molloy:1995rh}, let the average degree be the order parameter for a percolation transition. Suppose that the considered self-similar ensembles do have a non-zero percolation threshold at some critical value of the average degree $\langle k \rangle_c$. Consider a graph with the average degree below the threshold ($\langle k \rangle < \langle k \rangle_c$) which has no giant component. Since its subgraphs belong to the same ensemble, their percolation threshold is also $\langle k \rangle_c$. But since their average degree increases with their depth in the subgraph hierarchy, there exist deep enough subgraphs whose average degree is above the threshold ($\langle k \rangle_T > \langle k \rangle_c$). We thus arrive at a contradiction since a graph which does not have a giant component must contain subgraphs which do have giant components.

We next compute the size $g(p)$ of the giant component in bond percolations with bond occupation probability $p$, confirming the absence of the percolation threshold in the considered ensembles. We first focus on the equilibrium networks of types I and II, in which case bond percolation is equivalent to replacing the connection probability $r_{ij}$ with $p r_{ij}$. Given a node $i$ and a set of other nodes $\Upsilon$, the probability that $i$ is connected to at least one node in $\Upsilon$ is one minus the probability that $i$ is not connected to any node in $\Upsilon$, {\it i.e.}, $1-\exp{\left[\sum_{j\in \Upsilon}\ln{(1-pr_{ij})} \right]}$. Node $i$ belongs to the giant component if and only if it is connected to the giant component of the graph without $i$. If $\widetilde{g}_j(p)$ denotes the probability that this $i$-deprived component contains some other node $j$, then
\begin{equation}
g_i(p)=1-\exp{\left[\sum_{j \ne i}\widetilde{g}_j(p)\ln{(1-pr_{ij})}\right]}.
\label{g_i(p)}
\end{equation}
Since in small-world networks a single node cannot significantly affect the percolation properties of the rest of the graph, we identify $\widetilde{g}_j(p)=g_j(p)$, transforming Eq.~(\ref{g_i(p)}) into a self-consistent equation for $g_i(p)$. We note that Eq.~(\ref{g_i(p)}) does not use the treelike assumption. This equation is thus valid for the type II graphs with strong clustering as well as for zero-clustering type I graphs. It leads in the thermodynamic limit to the following expression for the probability $g(\kappa;p)$ that a node with expected degree $\kappa$ belongs to the giant component:
\begin{eqnarray}
g(\kappa ; p)&=&1-e^{-\kappa \psi(p)},\quad\text{where $\psi(p)$ satisfies}\\
\frac{\left[ \psi(p)\right]^{3-\gamma}}{a(p)}&=& \frac{\left[ \psi(p)\right]^{2-\gamma}}{\gamma-2}-\Gamma[2-\gamma,\psi(p)],\quad\text{with}\\
a_I(p)&=&\frac{(\gamma-2)^2}{\gamma-1} \langle k \rangle p,\\
a_{II}(p)&=&-\frac{(\gamma-2)^2}{\gamma-1} \frac{\langle k \rangle}{I} \int_0^{\infty} \ln{(1-p h(x))}dx
\end{eqnarray}
for types I and II, respectively. The size of the giant component is then
$
g(p)=\int \rho(\kappa)g(\kappa;p) d\kappa=1-(\gamma-1)E_{\gamma}[\psi(p)],
\label{g(p)}
$
where $E_{\gamma}$ is the extended exponential integral. In diluted networks with $p\ll1$, $a_I(p) \approx a_{II}(p)$, and the giant component size for both classes becomes
\begin{equation}
g(p) \sim \left[- \frac{(\gamma-2)^{\gamma-1}}{(\gamma-1)^{\gamma-2}\Gamma(2-\gamma)}   \langle k \rangle p \right]^{\frac{1}{3-\gamma}}.
\end{equation}
The value of the critical exponent $\beta$ in $g(p) \sim p^\beta$ is thus $\beta=1/(3-\gamma)$, agreeing with \cite{Cohen:2002jx}. We emphasize that in our case, this result is obtained without using the treelike assumption. Therefore, quite surprisingly, this exponent characterizes equilibrium scale-free networks with arbitrary clustering and degree correlations.
\begin{figure}[t]
\centerline{\includegraphics[width=3in]{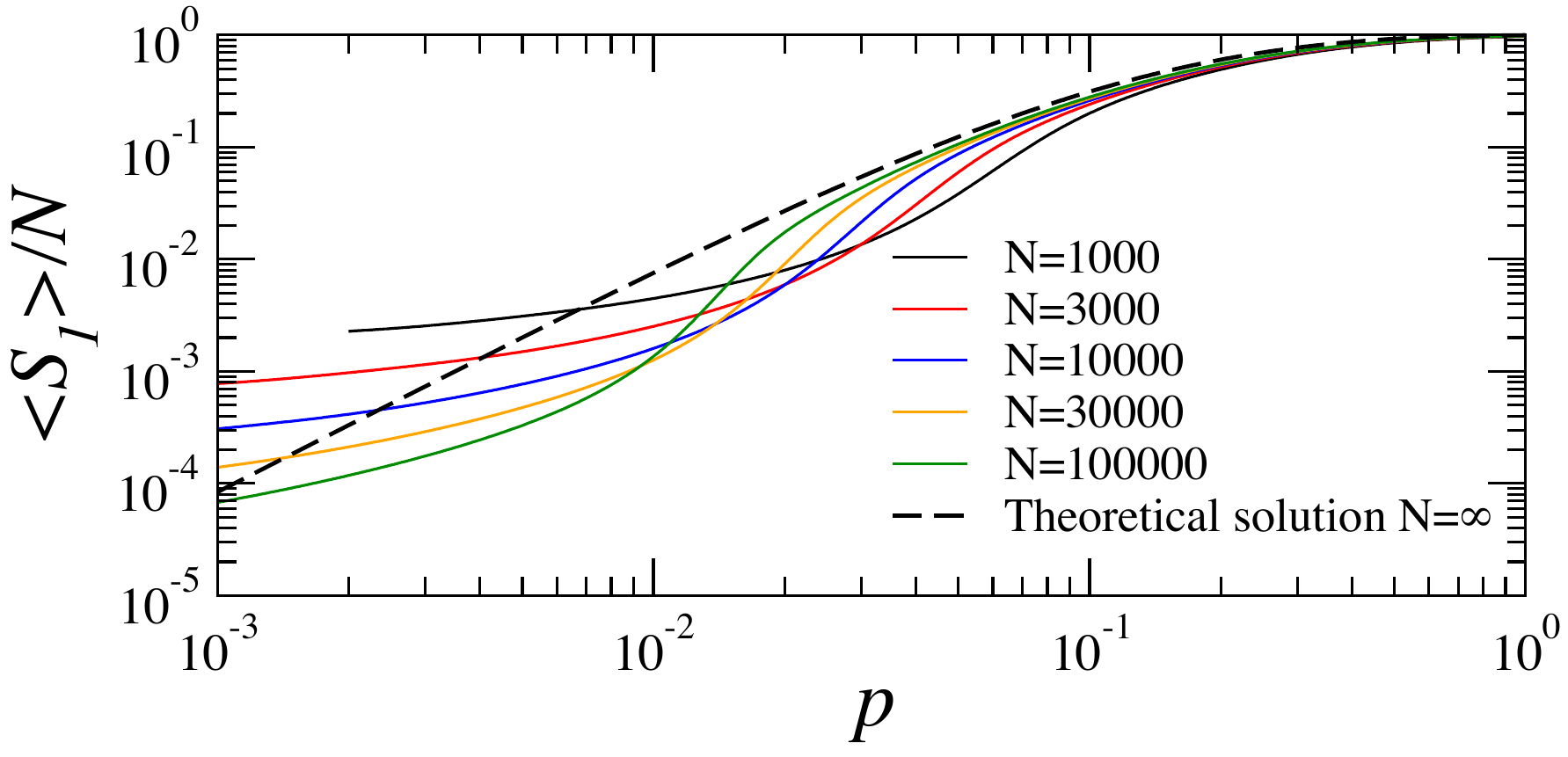}}
\caption{Relative size of the largest cluster $\langle S_1 \rangle /N$ vs.\ the analytical solution for type II networks with $\gamma=2.5$, $\langle k \rangle=3(\gamma-1)/(\gamma-2)$, $h(x)=e^{-x}$, and average clustering coefficient (measured over degrees larger than $k=1$) $\bar{c}=0.5$.
}
\label{fig2}
\vspace{-0.5cm}
\end{figure}

\begin{figure}[t]
\centerline{\includegraphics[width=3in]{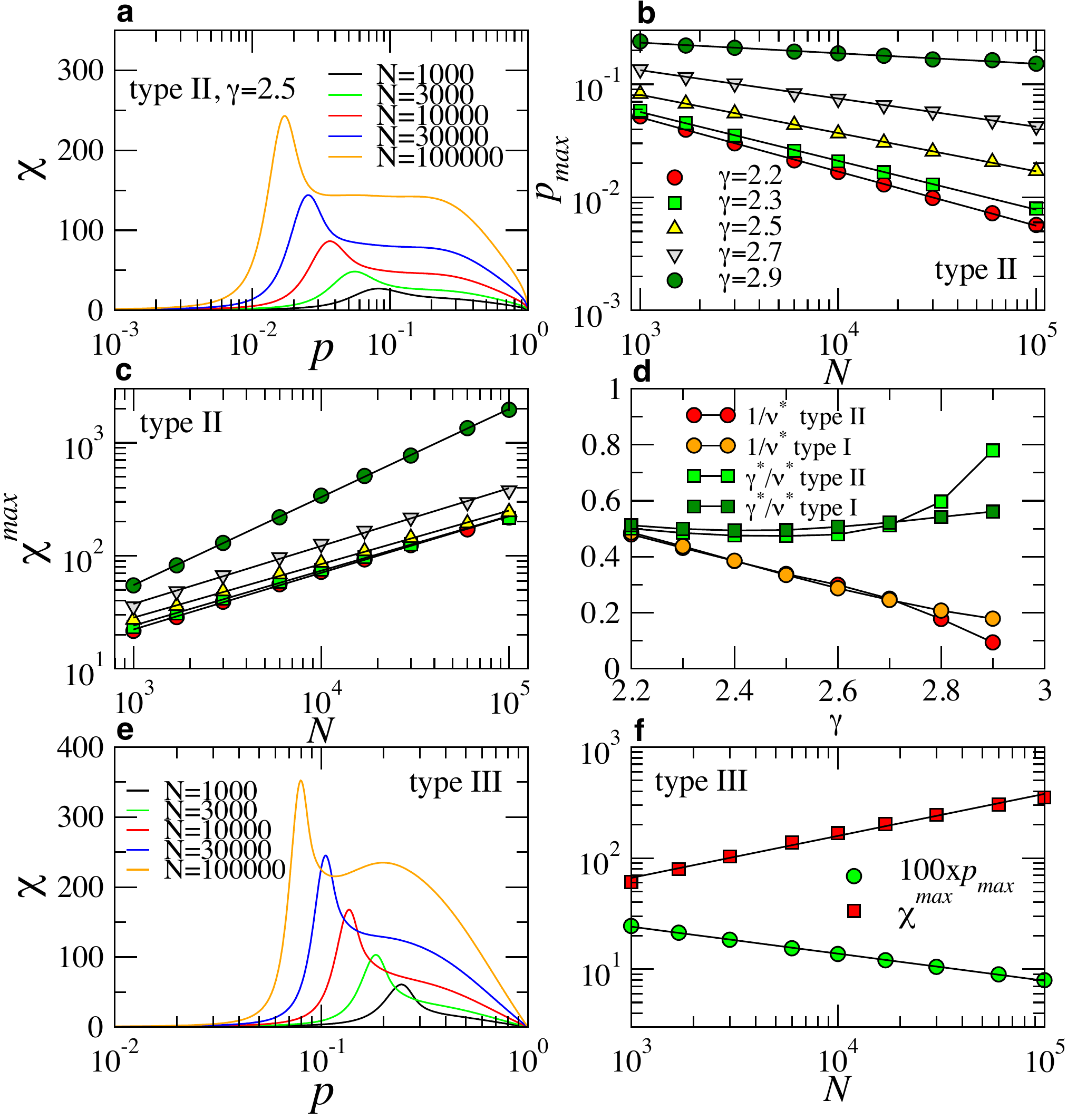}}
\caption{Bond percolation simulations for equilibrium (types I and II) and growing (type III) networks . {\bf a:} susceptibility $\chi$ as a function of bond occupation probability $p$ and graph size $N$ for the same network as in Fig.~\ref{fig2}. {\bf b} and {\bf c:} position $p_{max}$ and height $\chi^{max}$ of the peak of $\chi$ as functions of network size $N$. The straight lines are power law fits. {\bf d:} exponents $1/\nu^*$ and $\gamma^*/\nu^*$ in $p_{max}(N)\sim N^{-1/\nu^*}$ and $\chi^{max}(N) \sim N^{\gamma^*/\nu^*}$ for the type I and II graphs. {\bf e} and {\bf f:} Bond percolation simulations for non-equilibrium networks (type III) with $\eta=1/4$ ($\gamma=5$) and $m_0=2$. The measured values of the scaling exponents are $1/\nu^*=0.24(3)$ and $\gamma^*/\nu^*=0.3(8)$.
} \label{fig3}
\vspace{-0.5cm}
\end{figure}

In non-equilibrium networks of type III, we can compute an upper bound for $\beta$. Self-similarity of these networks, coupled with the observation that any node belonging to the giant component of a self-similar subgraph of a type III graph belongs also to the giant component of the graph itself, leads to inequality
\begin{equation}
g(p) \ge \frac{N_T}{N} g\left(\left[\frac{N}{N_T} \right]^{\eta}p \right).
\end{equation}
By choosing $N_T/N = p^{1/\eta}$, we obtain $g(p) \ge p^{1/\eta} g(1)$. Therefore the exponent $\beta$ satisfies $\beta \le 1/\eta=\gamma-1$. We see that growth reduces significantly this exponent, compared to the equilibrium case with the same $\gamma$.

We next check our analytic results against large-scale simulations. We generate type I and II networks using the connection probabilities in Eq.~(\ref{r_typeI}) and $h(x)=e^{-x}$, respectively. We do not allow $\kappa$'s above the natural cutoff $\kappa_c = N^{1/(\gamma-1)}$. For all the three graph types, for each graph size $N$ ranging from $10^3$ to $10^5$, and for each value of the bond occupation probability $p$, we generate $10^3$ graphs, and for each graph we perform bond percolation $10^4$ times. For each percolation we measure the size $S_1$ of the largest connected component in the graph using the fast algorithm of Newman and Ziff~\cite{Newman:2000fk}, and calculate the average $\langle S_1 \rangle$ of the largest component size and its fluctuations, i.e., susceptibility $\chi=\sqrt{\langle(S_1-\langle S_1 \rangle)^2\rangle}$, for each combination of $p$, $N$, and graph type.

Since the convergence to the thermodynamic limit in scale-free networks is slow~\cite{Boguna:2009pi}, it is difficult to accurately measure exponent $\beta$ in simulations. Nevertheless we observe an agreement, albeit slowly converging, between the analytical solution for $g(p)$ and simulations in Fig.~\ref{fig2}. In Fig.~\ref{fig3} we also show susceptibility $\chi(p,N)$ for equilibrium and growing networks. Susceptibility displays peaks whose positions $p_{max}(N)$ and heights $\chi^{max}(N)$
depend as power laws on the system size, $p_{max}(N)\sim N^{-1/\nu^*}$ and $\chi^{max}(N) \sim N^{\gamma^*/\nu^*}$. Taken together, these two results confirm that the giant component emerges at $p = p_{max}$, and that the percolation threshold vanishes in the thermodynamic limit $N\to\infty$ where $p_{max}\to0$ and $\chi^{max}\to\infty$.

In short, self-similar networks with subgraphs of growing average degree have no percolation threshold. The proof can be generalized to any processes with phase transitions whose critical points depend monotonously on the average degree. Examples include, among others, the absence of an epidemic threshold in epidemic spreading processes, or the absence of a paramagnetic phase in the Ising model on scale-free networks~\cite{Dorogovtsev:2008kx}.

The identification of percolation universality classes for general random networks is a notoriously difficult problem---details tend to prevail. Nevertheless, the results presented here lead us to conjecture that {\em self-similar\/} networks can be split into three general percolation universality classes, depending only on whether the average degree in the nested subgraph hierarchy increases, remains constant, or decreases with the subgraph depth, and independent of any other network properties, such as clustering, correlations, equilibrium vs.\ non-equilibrium classification, etc.

\begin{acknowledgments}
This work was supported by DGES Grant No.\ FIS2010-21781-C02-02; Generalitat de
Catalunya grant No.\ 2009SGR838; the Ram\'on y Cajal program of the Spanish Ministry of Science; MICINN Project No.\ BFU2010-21847-C02-02;
NSF Grants No.\ CNS-1039646, CNS-0964236 and CNS-0722070; DHS Grant No.\ N66001-08-C-2029;
and by Cisco Systems.
\end{acknowledgments}


\begin{thebibliography}{24}
\expandafter\ifx\csname natexlab\endcsname\relax\def\natexlab#1{#1}\fi
\expandafter\ifx\csname bibnamefont\endcsname\relax
  \def\bibnamefont#1{#1}\fi
\expandafter\ifx\csname bibfnamefont\endcsname\relax
  \def\bibfnamefont#1{#1}\fi
\expandafter\ifx\csname citenamefont\endcsname\relax
  \def\citenamefont#1{#1}\fi
\expandafter\ifx\csname url\endcsname\relax
  \def\url#1{\texttt{#1}}\fi
\expandafter\ifx\csname urlprefix\endcsname\relax\def\urlprefix{URL }\fi
\providecommand{\bibinfo}[2]{#2}
\providecommand{\eprint}[2][]{\url{#2}}

\bibitem[{\citenamefont{Smirnov}(2001)}]{Smirnov:FieldsMedal}
\bibinfo{author}{\bibfnamefont{S.}~\bibnamefont{Smirnov}}, \bibinfo{journal}{C.
  R. Acad. Sci. Paris Sr. I Math} \textbf{\bibinfo{volume}{333}},
  \bibinfo{pages}{239} (\bibinfo{year}{2001}).

\bibitem[{\citenamefont{Cardy}(2003)}]{Cardy:ExplainsStuff}
\bibinfo{author}{\bibfnamefont{J.}~\bibnamefont{Cardy}}, \bibinfo{journal}{Ann.
  Henri Poincar\'e} \textbf{\bibinfo{volume}{4, Suppl. 1}},
  \bibinfo{pages}{S371} (\bibinfo{year}{2003}).

\bibitem[{\citenamefont{Dorogovtsev et~al.}(2008)\citenamefont{Dorogovtsev,
  Goltsev, and Mendes}}]{Dorogovtsev:2008kx}
\bibinfo{author}{\bibfnamefont{S.~N.} \bibnamefont{Dorogovtsev}},
  \bibinfo{author}{\bibfnamefont{A.~V.} \bibnamefont{Goltsev}},
  \bibnamefont{and} \bibinfo{author}{\bibfnamefont{J.~F.~F.}
  \bibnamefont{Mendes}}, \bibinfo{journal}{Rev. Mod. Phys.}
  \textbf{\bibinfo{volume}{80}}, \bibinfo{pages}{1275} (\bibinfo{year}{2008}).

\bibitem[{\citenamefont{Molloy and Reed}(1995)}]{Molloy:1995rh}
\bibinfo{author}{\bibfnamefont{M.}~\bibnamefont{Molloy}} \bibnamefont{and}
  \bibinfo{author}{\bibfnamefont{B.}~\bibnamefont{Reed}},
  \bibinfo{journal}{Random Structures and Algorithms}
  \textbf{\bibinfo{volume}{6}}, \bibinfo{pages}{161} (\bibinfo{year}{1995}).

\bibitem[{\citenamefont{Cohen et~al.}(2000)\citenamefont{Cohen, Erez, ben
  Avraham, and Havlin}}]{Cohen:2000ss}
\bibinfo{author}{\bibfnamefont{R.}~\bibnamefont{Cohen}},
  \bibinfo{author}{\bibfnamefont{K.}~\bibnamefont{Erez}},
  \bibinfo{author}{\bibfnamefont{D.}~\bibnamefont{ben Avraham}},
  \bibnamefont{and} \bibinfo{author}{\bibfnamefont{S.}~\bibnamefont{Havlin}},
  \bibinfo{journal}{Phys. Rev. Lett.} \textbf{\bibinfo{volume}{85}},
  \bibinfo{pages}{4626} (\bibinfo{year}{2000}).

\bibitem[{\citenamefont{Krapivsky and Derrida}(2004)}]{Krapivsky:2004bc}
\bibinfo{author}{\bibfnamefont{P.~L.} \bibnamefont{Krapivsky}}
  \bibnamefont{and} \bibinfo{author}{\bibfnamefont{B.}~\bibnamefont{Derrida}},
  \bibinfo{journal}{Physica A} \textbf{\bibinfo{volume}{340}},
  \bibinfo{pages}{714} (\bibinfo{year}{2004}).

\bibitem[{\citenamefont{V\'{a}zquez and Moreno}(2003)}]{Vazquez:2003lg}
\bibinfo{author}{\bibfnamefont{A.}~\bibnamefont{V\'{a}zquez}} \bibnamefont{and}
  \bibinfo{author}{\bibfnamefont{Y.}~\bibnamefont{Moreno}},
  \bibinfo{journal}{Physical Review E} \textbf{\bibinfo{volume}{67}},
  \bibinfo{pages}{015101(R)} (\bibinfo{year}{2003});
\bibinfo{author}{\bibfnamefont{A.~V.} \bibnamefont{Goltsev}},
  \bibinfo{author}{\bibfnamefont{S.~N.} \bibnamefont{Dorogovtsev}},
  \bibnamefont{and} \bibinfo{author}{\bibfnamefont{J.~F.~F.}
  \bibnamefont{Mendes}}, \bibinfo{journal}{Phys. Rev. E}
  \textbf{\bibinfo{volume}{78}}, \bibinfo{pages}{051105}
  (\bibinfo{year}{2008}).

\bibitem[{\citenamefont{Serrano and
  Bogu{\~n}\'{a}}(2006{\natexlab{a}})}]{Serrano:2006qj}
\bibinfo{author}{\bibfnamefont{M.~A.} \bibnamefont{Serrano}} \bibnamefont{and}
  \bibinfo{author}{\bibfnamefont{M.}~\bibnamefont{Bogu{\~n}\'{a}}},
  \bibinfo{journal}{Phys. Rev. E} \textbf{\bibinfo{volume}{74}},
  \bibinfo{pages}{056114} (\bibinfo{year}{2006}{\natexlab{a}}).

\bibitem[{\citenamefont{Newman}(2003)}]{Newman:2003ft}
\bibinfo{author}{\bibfnamefont{M.~E.~J.} \bibnamefont{Newman}},
  \bibinfo{journal}{Phys Rev E} \textbf{\bibinfo{volume}{68}},
  \bibinfo{pages}{026121} (\bibinfo{year}{2003});
\bibinfo{author}{\bibfnamefont{M.~A.} \bibnamefont{Serrano}} \bibnamefont{and}
  \bibinfo{author}{\bibfnamefont{M.}~\bibnamefont{Bogu{\~n}\'{a}}},
  \bibinfo{journal}{Phys Rev E} \textbf{\bibinfo{volume}{74}},
  \bibinfo{pages}{056115} (\bibinfo{year}{2006}{\natexlab{b}});
\bibinfo{author}{\bibfnamefont{M.~A.} \bibnamefont{Serrano}} \bibnamefont{and}
  \bibinfo{author}{\bibfnamefont{M.}~\bibnamefont{Bogu{\~n}\'{a}}},
  \bibinfo{journal}{Phys. Rev. Lett.} \textbf{\bibinfo{volume}{97}},
  \bibinfo{pages}{088701} (\bibinfo{year}{2006}{\natexlab{c}});
\bibinfo{author}{\bibfnamefont{M.~E.~J.} \bibnamefont{Newman}},
  \bibinfo{journal}{Phys. Rev. Lett.} \textbf{\bibinfo{volume}{103}},
  \bibinfo{pages}{058701} (\bibinfo{year}{2009});
\bibinfo{author}{\bibfnamefont{J.~P.} \bibnamefont{Gleeson}},
  \bibinfo{journal}{Phys. Rev. E} \textbf{\bibinfo{volume}{80}},
  \bibinfo{pages}{036107} (\bibinfo{year}{2009}).

\bibitem[{\citenamefont{Bogu{\~{n}}\'{a} and
  Pastor-Satorras}(2003)}]{Boguna:2003ja}
\bibinfo{author}{\bibfnamefont{M.}~\bibnamefont{Bogu{\~{n}}\'{a}}}
  \bibnamefont{and}
  \bibinfo{author}{\bibfnamefont{R.}~\bibnamefont{Pastor-Satorras}},
  \bibinfo{journal}{Phys Rev E} \textbf{\bibinfo{volume}{68}},
  \bibinfo{pages}{036112} (\bibinfo{year}{2003}).

\bibitem[{\citenamefont{Park and Newman}(2003)}]{Park:2003qy}
\bibinfo{author}{\bibfnamefont{J.}~\bibnamefont{Park}} \bibnamefont{and}
  \bibinfo{author}{\bibfnamefont{M.~E.~J.} \bibnamefont{Newman}},
  \bibinfo{journal}{Phys Rev E} \textbf{\bibinfo{volume}{68}},
  \bibinfo{pages}{026112} (\bibinfo{year}{2003});
\bibinfo{author}{\bibfnamefont{D.}~\bibnamefont{Garlaschelli}}
  \bibnamefont{and} \bibinfo{author}{\bibfnamefont{M.~I.}
  \bibnamefont{Loffredo}}, \bibinfo{journal}{Phys. Rev. E}
  \textbf{\bibinfo{volume}{78}}, \bibinfo{pages}{015101}
  (\bibinfo{year}{2008}).

\bibitem[{\citenamefont{Bogu{\~n}\'{a}
  et~al.}(2004)\citenamefont{Bogu{\~n}\'{a}, Pastor-Satorras, and
  Vespignani}}]{Boguna:2004eh}
\bibinfo{author}{\bibfnamefont{M.}~\bibnamefont{Bogu{\~n}\'{a}}},
  \bibinfo{author}{\bibfnamefont{R.}~\bibnamefont{Pastor-Satorras}},
  \bibnamefont{and}
  \bibinfo{author}{\bibfnamefont{A.}~\bibnamefont{Vespignani}},
  \bibinfo{journal}{Eur. Phys. J. B} \textbf{\bibinfo{volume}{38}},
  \bibinfo{pages}{205} (\bibinfo{year}{2004}).

\bibitem[{\citenamefont{Serrano et~al.}(2008)\citenamefont{Serrano, Krioukov,
  and Bogu{\~{n}}\'{a}}}]{Serrano:2008ga}
\bibinfo{author}{\bibfnamefont{M.~{\'A}.} \bibnamefont{Serrano}},
  \bibinfo{author}{\bibfnamefont{D.}~\bibnamefont{Krioukov}}, \bibnamefont{and}
  \bibinfo{author}{\bibfnamefont{M.}~\bibnamefont{Bogu{\~{n}}\'{a}}},
  \bibinfo{journal}{Phys Rev Lett} \textbf{\bibinfo{volume}{100}},
  \bibinfo{pages}{078701} (\bibinfo{year}{2008}).

\bibitem[{\citenamefont{Alvarez-Hamelin
  et~al.}(2008)\citenamefont{Alvarez-Hamelin, Dall'Asta, Barrat, and
  Vespignani}}]{Alvarez-Hamelin:2008fv}
\bibinfo{author}{\bibfnamefont{J.~I.} \bibnamefont{Alvarez-Hamelin}},
  \bibinfo{author}{\bibfnamefont{L.}~\bibnamefont{Dall'Asta}},
  \bibinfo{author}{\bibfnamefont{A.}~\bibnamefont{Barrat}}, \bibnamefont{and}
  \bibinfo{author}{\bibfnamefont{A.}~\bibnamefont{Vespignani}},
  \bibinfo{journal}{Networks and Heterogeneous Media}
  \textbf{\bibinfo{volume}{3}}, \bibinfo{pages}{371} (\bibinfo{year}{2008}).

\bibitem[{\citenamefont{Cohen et~al.}(2002)\citenamefont{Cohen, ben Avraham,
  and Havlin}}]{Cohen:2002jx}
\bibinfo{author}{\bibfnamefont{R.}~\bibnamefont{Cohen}},
  \bibinfo{author}{\bibfnamefont{D.}~\bibnamefont{ben Avraham}},
  \bibnamefont{and} \bibinfo{author}{\bibfnamefont{S.}~\bibnamefont{Havlin}},
  \bibinfo{journal}{Phys. Rev. E} \textbf{\bibinfo{volume}{66}},
  \bibinfo{pages}{036113} (\bibinfo{year}{2002});
\bibinfo{author}{\bibfnamefont{M.} \bibnamefont{Ostilli}}
  \bibnamefont{and} \bibinfo{author}{\bibfnamefont{J.~F.~F.} \bibnamefont{Mendes}},
  \bibinfo{journal}{Eur. Phys. Lett.} \textbf{\bibinfo{volume}{92}},
  \bibinfo{pages}{40013} (\bibinfo{year}{2010}).

\bibitem[{\citenamefont{Newman and Ziff}(2000)}]{Newman:2000fk}
\bibinfo{author}{\bibfnamefont{M.~E.~J.} \bibnamefont{Newman}}
  \bibnamefont{and} \bibinfo{author}{\bibfnamefont{R.~M.} \bibnamefont{Ziff}},
  \bibinfo{journal}{Phys. Rev. Lett.} \textbf{\bibinfo{volume}{85}},
  \bibinfo{pages}{4104} (\bibinfo{year}{2000}).

\bibitem[{\citenamefont{Bogu\~n\'a et~al.}(2009)\citenamefont{Bogu\~n\'a,
  Castellano, and Pastor-Satorras}}]{Boguna:2009pi}
\bibinfo{author}{\bibfnamefont{M.}~\bibnamefont{Bogu\~n\'a}},
  \bibinfo{author}{\bibfnamefont{C.}~\bibnamefont{Castellano}},
  \bibnamefont{and}
  \bibinfo{author}{\bibfnamefont{R.}~\bibnamefont{Pastor-Satorras}},
  \bibinfo{journal}{Phys. Rev. E} \textbf{\bibinfo{volume}{79}},
  \bibinfo{pages}{036110} (\bibinfo{year}{2009}).

\end{thebibliography}

\end{document}